# $NaRe_2(PO_4)_3$ phosphate-based ceramic with kosnarite structure as a matrix for technetium immobilization. Production. Properties

**L.S. Alekseeva**[(*),1], **A.V. Nokhrin**[(**)1], **A.I. Orlova**[1], **M.S. Boldin**[1], **E.A. Lantcev**[1], **A.A. Murashov**[1], **K.K. Korchenkin**[2], **D.V. Ryabkov**[2], **V.N. Chuvil'deev**[1]

[1] Lobachevsky State University of Nizhny Novgorod, Russian Federation, 603022 Nizhny Novgorod, 23 Gagarina ave.

[2] V.G. Khlopin Radium Institute, JSC, Russian Federation, 194021 St. Petersburg, 28 2nd Murinsky ave.

### Abstract

$NaRe_2(PO_4)_3$ phosphate-based ceramic with the structure of kosnarite mineral was obtained by spark plasma sintering. Rhenium (Re) served as a chemical and structural analog of technetium. The ceramic relative density was 85%. The mechanism of Re static leaching from $NaRe_2(PO_4)_3$ ceramic at room temperature was investigated. The leaching rate of rhenium was $1.3 \times 10^{-5}$ g/(cm$^2$×day).

### Introduction

Technetium-99 ($^{99}$Tc) is one of the most widespread and long-lived radiotoxic isotopes in spent nuclear fuel (SNF). It is one of the key elements in SNF separation strategies such as UREX + for isolation and encapsulation in solid waste forms [1]. Immobilizing $^{99}$Tc is a complex scientific and practical challenge. Thus far, immobilization of $^{99}$Tc as metal alloys [2-4], cements [5-7], and glasses [2, 8-9] has been researched. During glass transition, which is the most widely used process for high-level radioactive waste (RAW), technetium is partly oxidized or disproportionated and therefore evaporates as heptaoxide, causing environmental pollution. An alternative approach to immobilizing $^{99}$Tc consists in introducing a cation into a mineral-like matrix [10-16], which has the potential to help avoid formation of highly mobile pertechnetate $TcO_4^-$ ions during RAW processing.

---

[(*)] corresponding author 1 (golovkina_lyudmila@mail.ru)

[(**)] corresponding author 2 (nokhrin@nifti.unn.ru)

Tc-containing compounds with the structures of pyrochlore ($Nd_2Tc_2O_7$), perovskite ($SrTcO_3$) and lamellar perovksite ($Sr_2TcO_4$) have been studied in [13, 14]. 28 days of hydrolytic tests of $Nd_2Tc_2O_7$ pyrochlore-based ceramic helped establish the Tc leaching rate to be $1.48·10^{-7}$ g/(mm$^2$·day), which is approximately four times slower than the rate of leaching from borosilicate glass (~$6.43·10^{-7}$ g/(mm$^2$·day)). A disadvantage of the compounds studied in [13, 14] is the long (2–10 days) duration of high-temperature annealing during solid-phase synthesis.

Including technetium in compounds in $Mg_2Ti_{1-x}Tc_xO_4$ spinels has been studied in [11, 16]. Synthesis of such compounds is demanding and time consuming and their leaching rate was $(3-7)·10^{-3}$ g/(m$^2$·day) after testing for 40 days at room temperature.

In [15] describes compounds with NZP structure of $ARe_2(PO_4)_3$ (where A = AM [alkali metal]). These compounds were produced by fusing concentrated phosphoric acid, ammonium perrhenate, and AM chlorides, further annealed at 400 °C in air and at 550 °C under an inert atmosphere in a vacuum-sealed ampoule. There have been no studies of chemical stability of the synthesized compounds.

This paper researches $NaRe_2(PO_4)_3$ phosphate with kosnarite structure as a potential ceramic matrix for immobilizing $^{99}$Tc. Rhenium (Re) served as a chemical and structural analog of technetium. Solution chemistry was used to obtain $NaRe_2(PO_4)_3$ while the ceramic was produced through Spark Plasma Sintering (SPS) [17].

**Materials and methods**

A solution of 1M orthophosphoric acid was gradually added to a mixture of sodium chloride and ammonium perrhenate solutions taken in stoichiometric amounts with constant vigorous stirring in order to obtain $NaRe_2(PO_4)_3$ powders with the NZP structure. Their synthesis occurred as per the following formula: $NaCl + 2NH_4ReO_4 + 3H_3PO_4 = NaRe_2(PO_4)_3 + HCl + 2NH_4OH + 4H_2O$. The resulting clear solution was evaporated at 80 °C with constant stirring, dried at 200 and 300 °C until moisture was completely removed and the intermediate reaction products partially decomposed. The load turned black after annealing. The resulting powder was annealed in a vacuum-sealed quartz ampoule at 600, 700, or 800 °C for 12 hours.

A Shimadzu LabX XRD-6000 X-ray diffractometer (CuK$_\alpha$ filtered radiation) was used to determine the phase composition of the powders and ceramics. A DSC-204 F1 Phoenix thermal analyzer was used to perform a differential thermal analysis within 25-1200 °C range. The heating rate was 10 °C/min. A Shimadzu FTIR-8400S IR Fourier spectrophotometer was used to study the functional composition of the compounds at room temperature within a frequency range of 400-4000 cm$^{-1}$.

A Dr. Sinter model SPS-625 was used to produce the ceramics. The powders were placed in a graphite mold with an inner diameter of 10 mm and heated by millisecond pulses of direct electric current of high power (up to 3 kA). A Chino IR-AH pyrometer focused on the graphite mold surface was used to measure the sintering temperature. Sintering was performed in vacuum (6 Pa). The accuracy of temperature measurements was ± 10 °C, with a pressure maintenance accuracy of 1 MPa. The dilatometer in the Dr. Sinter model SPS-625 was used to monitor the shrinkage and shrinkage rate of the powders.

A Sartorius CPA balance was used to measure the density of the sintered samples by hydrostatic weighing in distilled water. Microstructural parameters of the samples were analyzed using a Jeol JSM-6490 scanning electron microscope (SEM) with an Oxford Instruments INCA 350 X-ray microanalyzer.

The chemical stability of the ceramics was studied by leaching in a static mode during 28 days. The tests were performed at room temperature in distilled water. The Re concentration in water samples was determined using an ELEMENT 2 high-resolution inductively coupled plasma mass spectrometer using external calibration. Calibration was performed using the ICP-MS-68A-A High-Purity Standards solutions and iDplus Performance time-of-flight mass spectrometer.

During the experiment, the normalized weight loss was calculated using the following formula:

$$NL_i = a_{ki} / (M_{oi} \times S), \qquad (1)$$

where $NL_i$ is the normalized weight loss of element $i$, g/cm$^2$; $a_{ki}$ is the mass of component $i$ dissolved during leaching, g; $M_{0i}$ is the mass concentration of the element in the sample at the beginning of testing, g/g; $S$ is the sample surface area, cm$^2$.

The leaching rate $R_i$ was calculated using the following formula:

$$R_i = NL_i / t_n, \qquad (2)$$

where $t_n$ is the time interval, days.

The DeGroot – van der Sloot model [18] was used to determine the mechanism of cation leaching from the ceramic, which can be represented as the following equation:

$$\lg B_i = A \lg t + \text{const}, \qquad (3)$$

where $B_i$ is the total yield of Re from the sample during contact with water, mg/m$^2$; $t$ is the contact time, days. The $B_i$ value was calculated using the following formula:

$$B_i = C_i(L/S)\sqrt{t_n} / (\sqrt{t_n} - \sqrt{t_{n-1}}), \qquad (4)$$

where $C_i$ is the concentration of Re in the solution by the end of the $n^{th}$ period, mg/l; $L/S$ is the ratio of solution volume to sample surface area, l/m$^2$; $t_n$ and $t_{n-1}$ is the time of the $n$- and $n$-1 experiment stages, respectively, in days.

The following leaching mechanisms correspond to the values of the coefficient A in equation (3): <0.35 – leaching from the surface of the compound; 0.35–0.65 – diffusion from inner layers; >0.65 – dissolution of the compound surface layer [19, 20].

**Results and discussion**

According to the XRD data (Fig. 1), a monophase product is obtained after annealing at 700 °C. Once annealed at 600 and 800 °C, the powder contains insignificant traces of ReP$_2$O$_7$ rhenium pyrophosphate and ReO$_2$ rhenium (IV) oxide, respectively. The resulting monophasic compounds crystallized in the expected NZP structure and belonged to the R-3c space group.

Further studies were performed with the powder synthesized at 700 °C.

The particle size distribution of the powder was heterogeneous. The electron microscopy data suggested that the synthesized powder contained two types of particles: large faceted particles approx. 10 μm in size, which are likely to be single crystals, and agglomerates consisting of particles up to 1 μm in size (Fig. 2a, b). Energy dispersive microanalysis confirmed the presence of Re in the structure of the synthesized powder (Fig. 2c).

Differential scanning calorimetry (DSC) (Fig. 3) revealed an endothermic effect in the temperature range of 970-1030 °C, which corresponded to the decomposition of NaRe$_2$(PO$_4$)$_3$ and was confirmed by the thermal gravimetric analysis (TGA).

A wide band in the range of 980-1120 cm$^{-1}$ and a high-frequency band at 1247 cm$^{-1}$ of the IR spectrum under study (Fig. 4) can be attributed to asymmetric stretching $\nu_3$ vibrations of the PO$_4^{3-}$ ion. We believe that a low-intensity high-frequency band in the range of 1247 cm$^{-1}$, which is not typical for phosphates, may be associated with the influence of the Re$^{4+}$ highly charged ion that polarizes the P–O–Re bond. The band in the range of 883 cm$^{-1}$ corresponded to symmetric stretching $\nu_1$ vibrations. The bands in the 640-438 cm$^{-1}$ range belong to bending vibrations: the bands within 650-540 cm$^{-1}$ corresponded to asymmetric $\nu_4$ vibrations while the 445 см$^{-1}$ band corresponded to symmetric $\nu_2$ vibrations.

SPS of NaRe$_2$(PO$_4$)$_3$ powder was performed under constant heating rate. There was no isothermal holding at the sintering temperature. Table 1 contains the main sintering parameters of the ceramic samples – heating rate $V_h$, average applied uniaxial pressure P, sintering temperature $T_s$ and isothermal time $t_s$ at sintering temperature.

**Table 1.** Main parameters of sintering $NaRe_2(PO_4)_3$ powders, density and hydrolytic stability of the resulting samples

| Sequence | $T_s$, °C | $V_h$, °C/min | $t_s$, min | P, MPa | ρ, g/cm³ | $ρ_{rel}$, % | $NL_{Re}$, g/cm² | $R_{Re}$, g/(cm²·day) |
|---|---|---|---|---|---|---|---|---|
| 1 | 1100 | 50 | 0 | 70 | - | - | - | - |
| 2 | 800 | 50 | 0 | 70 | 4.154 | 84.9 | 4.4·10⁻⁴ | 1.3·10⁻⁵ |
| 3 | 800 | 100 | 0 | 70 | 4.174 | 85.3 | - | - |
| 4 | 800 | 200 | 0 | 70 | 4.168 | 85.2 | - | - |

Preliminary sintering of $NaRe_2(PO_4)_3$ powders was performed by heating them to 1100 °C at a rate of 50 °C/min (Seq. # 1) to define the optimal SPS modes. Fig. 5a shows the sintering modes as dependences between temperature (T, °C), applied uniaxial pressure (P, kN), vacuum levels (Vac, Pa), and SPS time. Fig. 5b shows the dependence between shrinkage (L, mm), shrinkage rate (S, mm/s), and the heating temperature. Fig. 5b demonstrates that the powder shrinkage starts at T = 600 °C; with the shrinkage rate being small and not exceeding $S_{max}$ ~ 4.5·10⁻³ mm/s (Fig. 5b). A shrinkage peak was observed at 1000 °C, which is likely to be associated with a phase transition in $NaRe_2(PO_4)_3$ powder. It should also be noted that at temperatures above 850 °C powder decomposition was observed, which manifested itself in a decrease in the vacuum level in the sintering chamber of the Dr. Sinter model SPS-625 (Fig. 5a). Sample decomposition during SPS upon heating to 1100 °C prevented its density from being reliably determined through the Archimedes' principle.

As a way to minimize or exclude decomposition of the powder sample, further sintering of the ceramics was performed at $T_s$ = 800 °C with different heating rates (Seq. # 2-4). It is of interest to note that the maximum shrinkage rates $S_{max}$ increased slightly from ~0.2·10⁻³ to ~1.5·10⁻³ mm/s with an increase in the heating rate from 50 to 200 °C/min. The sintered samples were a powder compressed to a density $ρ_{rel}$ ≈ 85% with a particle size similar to the size of the initial powder particles d = 1-10 μm (Fig. 6). After sintering, many particles retained their faceted shape, indicating a low intensity of diffusion during SPS. The post-sintering density was insufficient to produce durable ceramics. The phase composition of the ceramic did not change after sintering (Fig. 7).

Table 1 shows the minimal rates of Re leaching from $NaRe_2(PO_4)_3$ samples. Fig. 8 shows graphs of dependence between the normalized weight loss $NL_i$, leaching rate $R_i$, and testing time $t$. The resulting data suggest that the rhenium leaching rate on day 28 was R = $1.3 \cdot 10^{-5}$ g/(cm$^2$·day). The calculated leaching rate is slightly lower, yet comparable to the rate of technetium leaching from other mineral-like compounds [11, 13, 14, 16]. It should be noted that the real surface area $S$ of the ceramics under study is greater than the value calculated based on geometric dimensions because of their increased porosity (Fig.6). We therefore believe that these results characterize the terminal value R for this compound.

Using formula (3) to determine the mechanism of rhenium leaching from $NaRe_2(PO_4)_3$ ceramic, the dependence between coefficient $B$ and the experiment time $t$ was plotted in logarithmic coordinates (Fig. 9). The data in Fig. 9 show that the value of coefficient $A$ is –0.38-0.3. It allows us to conclude that rhenium leaching occurs through washing out of the open surface of $NaRe_2(PO_4)_3$ ceramic. This result is qualitatively in good agreement with the previous conclusion that open porosity has a significant effect on the high rate of Re leaching from $NaRe_2(PO_4)_3$ ceramic.

**Conclusions**

$NaRe_2(PO_4)_3$ monophasic phosphate with the structure of the kosnarite mineral was obtained by synthesis at 700 °C. An increased or decreased synthesis temperature does not allow a monophasic product to be obtained. The synthesized $NaRe_2(PO_4)_3$ compound manifests a non-uniform particle size distribution, with large single-crystal particles of ~ 10 μm and agglomerated μm particles present in the powder.

In terms of achieving the maximum possible density of the ceramic ($\rho_{rel}$= 85.3%) while maintaining the phase composition, the optimal SPS mode consists in heating the sample to $T_s$ = 800 °C at a rate of $V_h$ = 100 °C/min under uniaxial pressure of P = 70 MPa.

The terminal rate of rhenium leaching from $NaRe_2(PO_4)_3$ ceramic on Day 28 was $1.3 \cdot 10^{-5}$ g/(cm$^2$·day), which makes it possible to class the phosphate under study as highly hydrolytically stable. It was established that the dominant mechanism of rhenium leaching in the static mode at room temperature is the leaching of cations from the surface of the ceramic.

**Acknowledgements**

This work has been performed with the support of Research and Educational Center of Nizhny Novgorod region in the framework of the agreement No 16-11-2021/49 (contract reference number 000000S407521QRR0002) and under contract No. 217/4571-D "Validation of radiation resistance of matrices. Validation of radiation resistance of phosphate ceramic matrices. Technology testing support" between the Lobachevsky State University of Nizhny Novgorod (UNN) and V.G. Khlopin Radium Institute.

**List of figures**

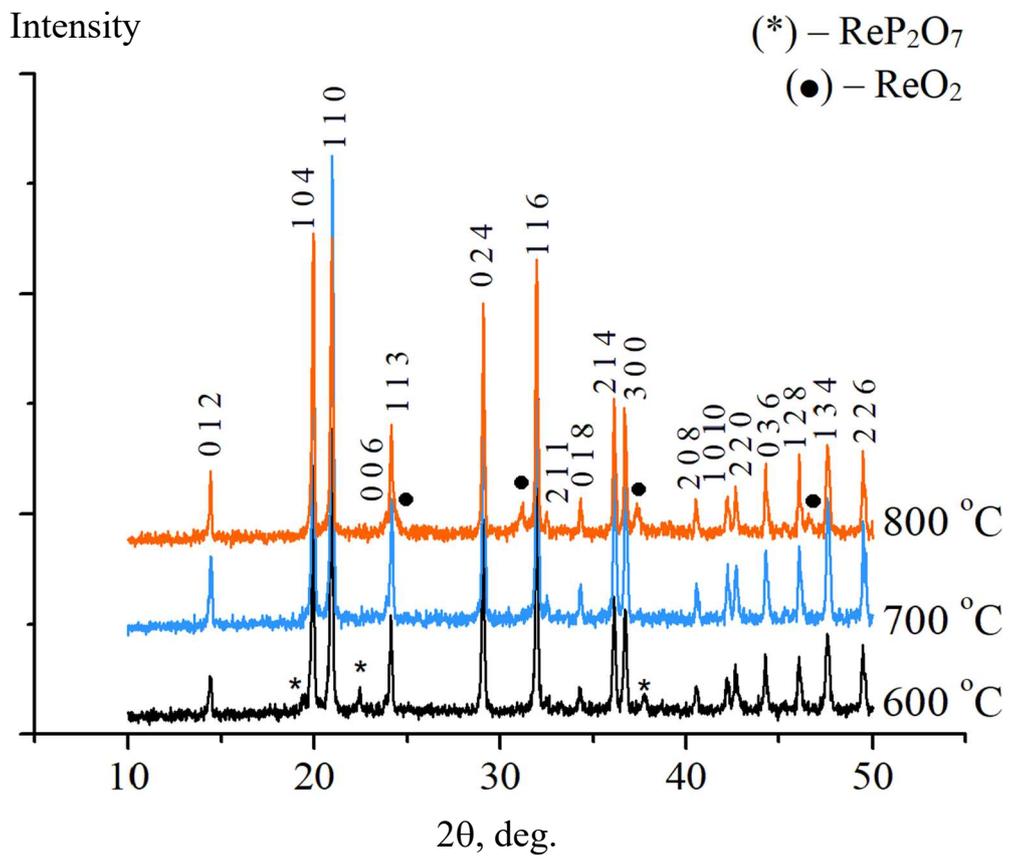

Figure 1

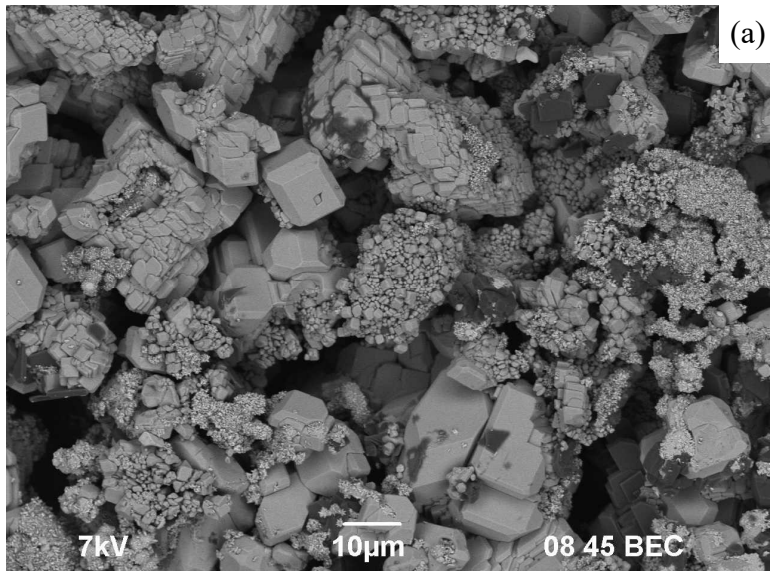
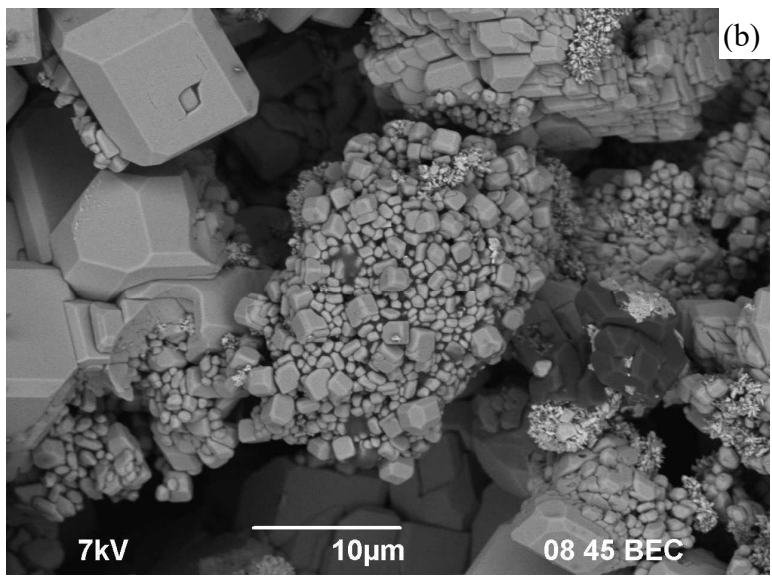
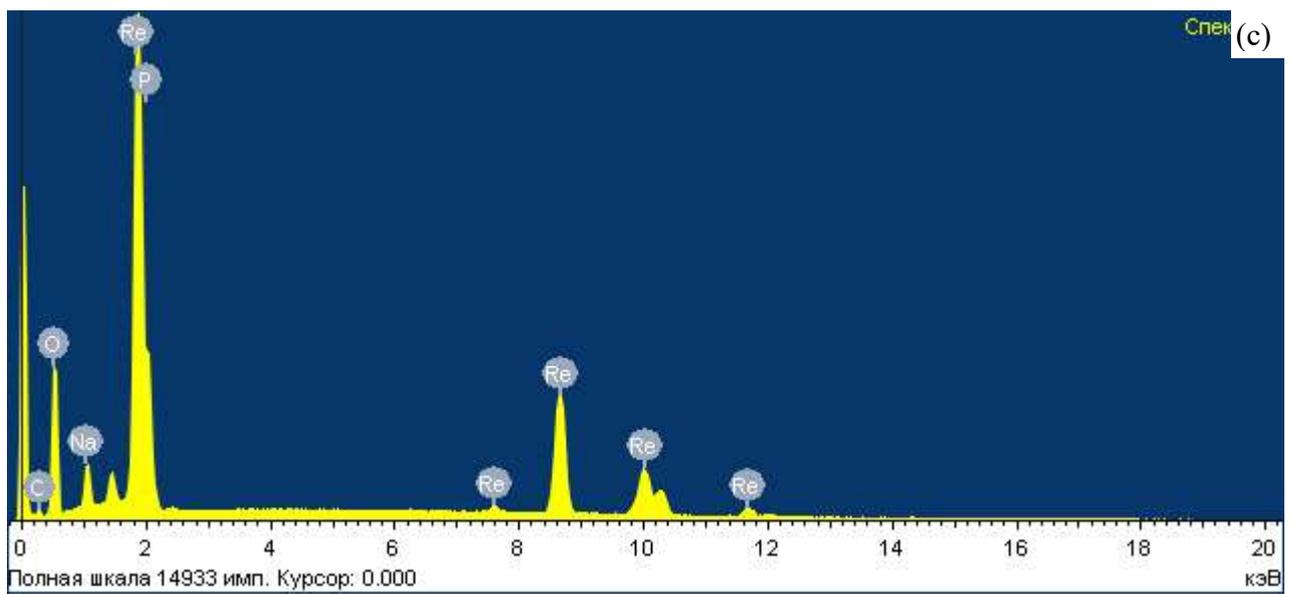

Figure 2

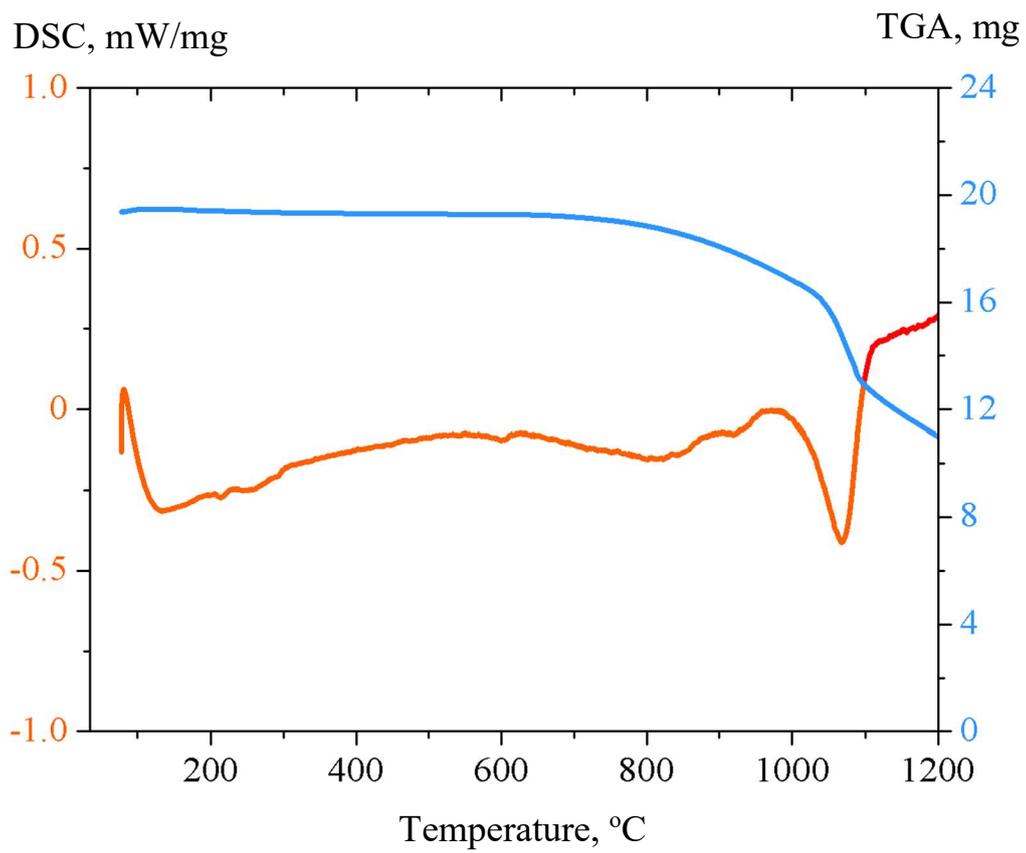

Figure 3

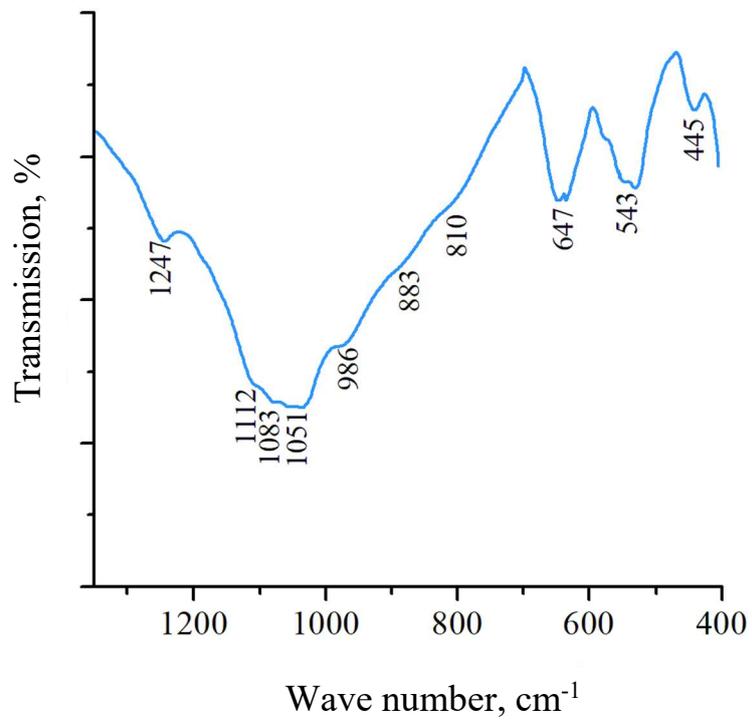

Figure 4

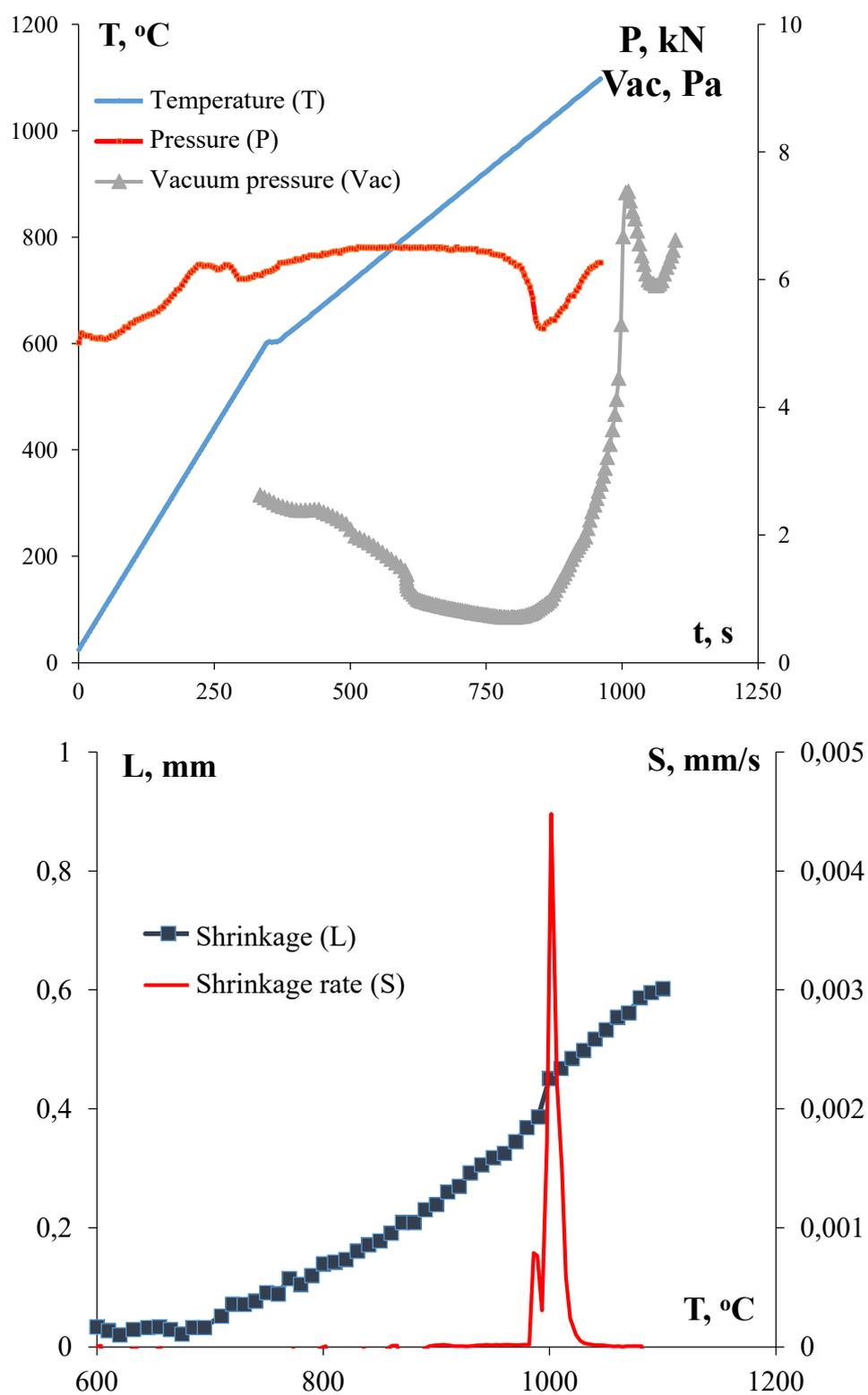

Figure 5

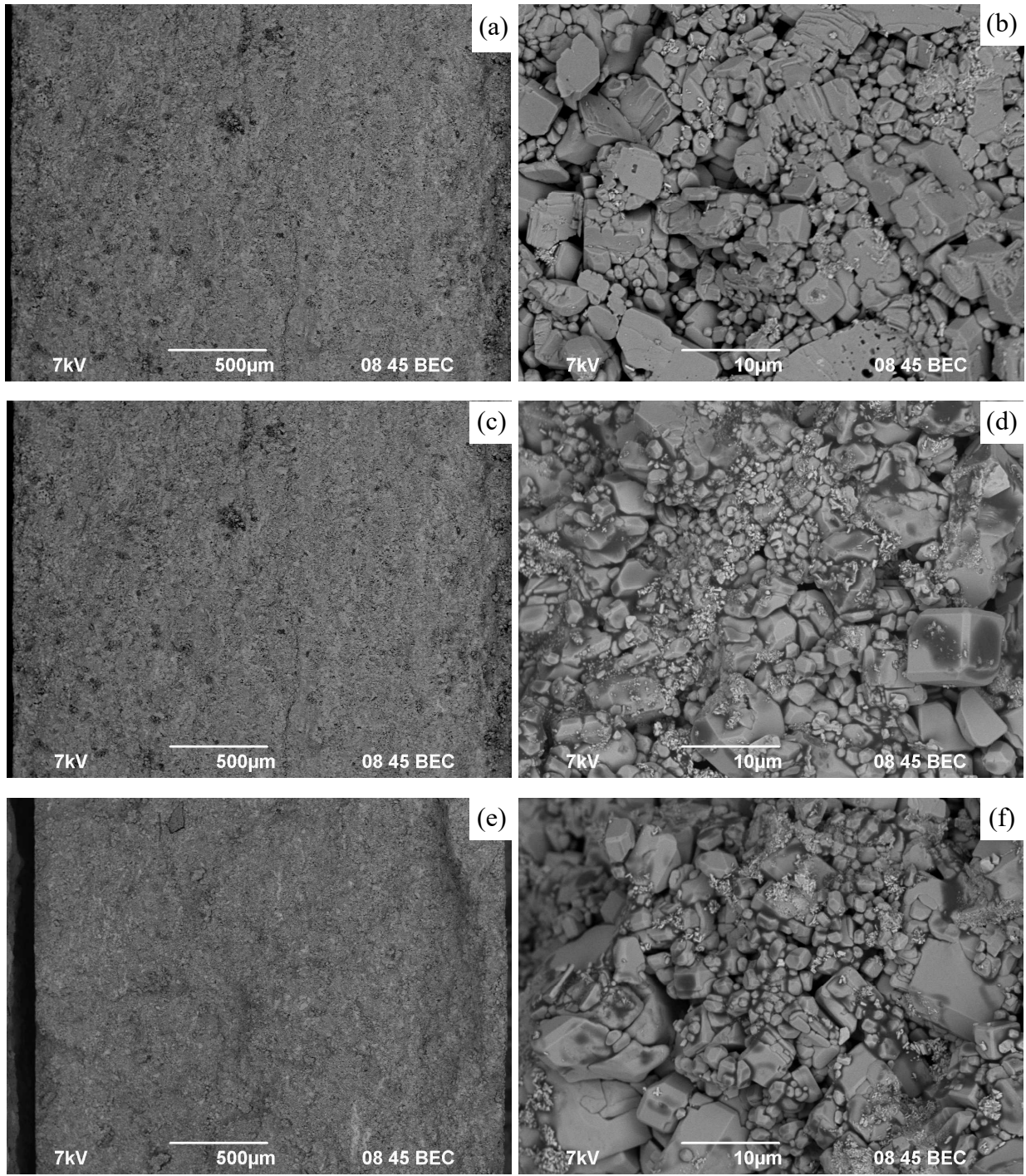

Figure 6

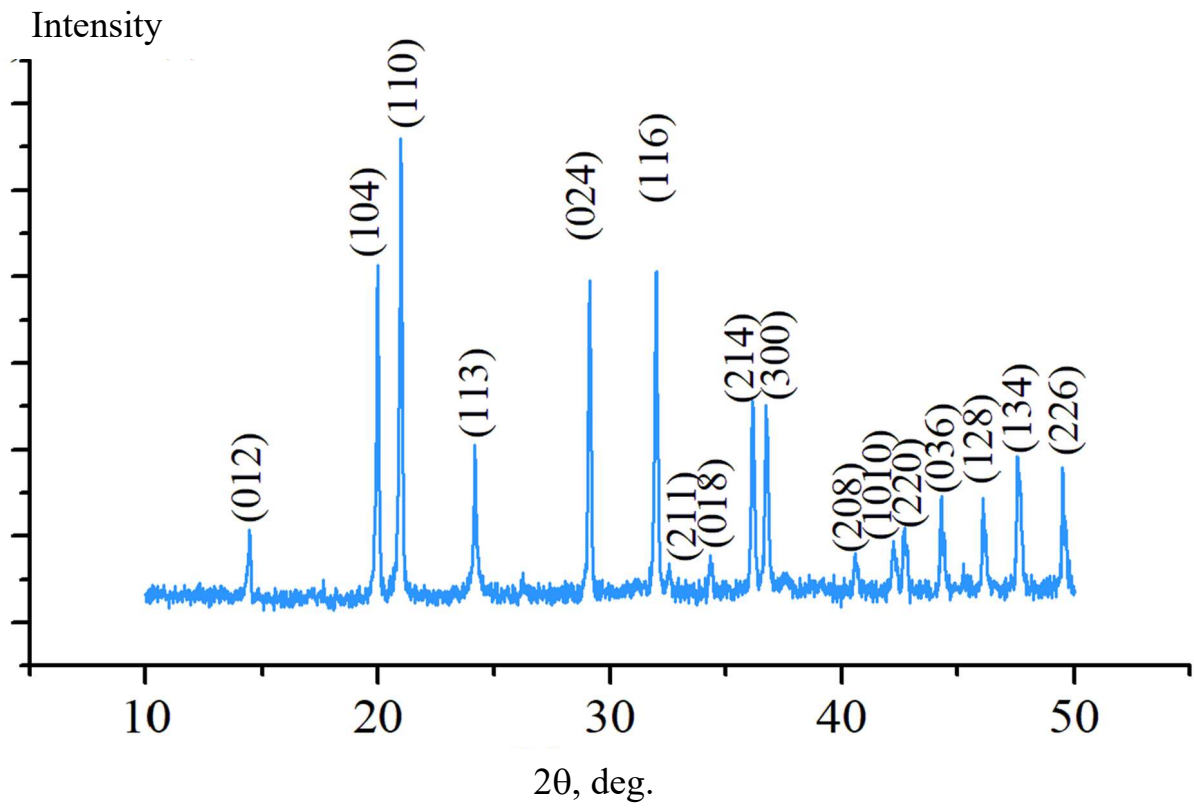

Figure 7

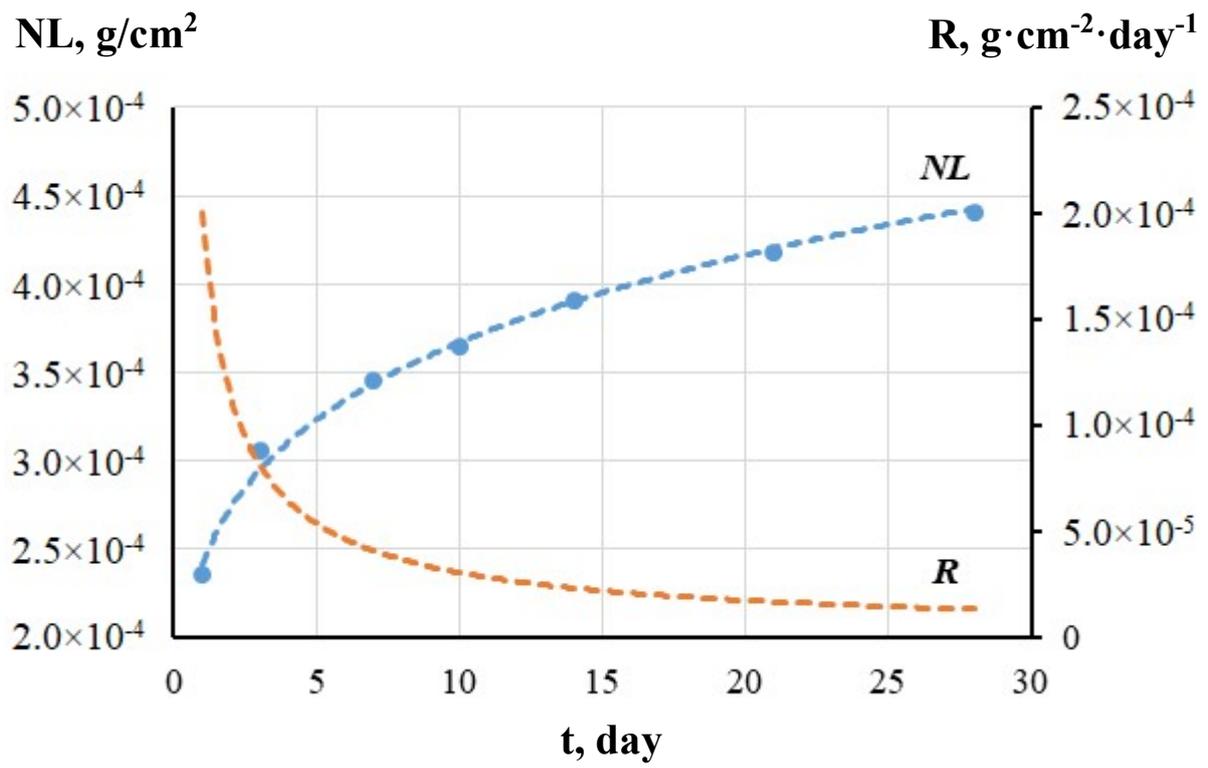

Figure 8

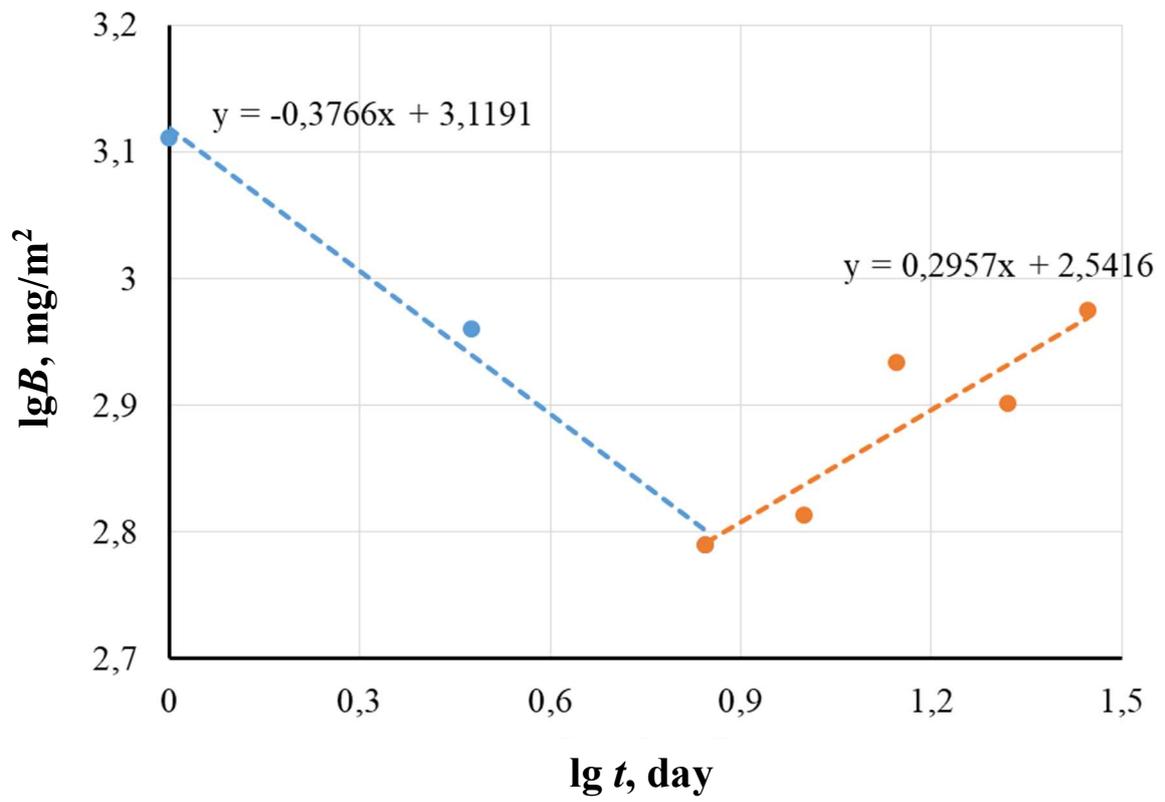

Figure 9